\documentclass[preprint]{vgtc}               

\graphicspath{{figures/}{pictures/}{images/}{./}} 

\usepackage{times}                     

\usepackage{tabu}                      
\usepackage{booktabs}                  
\usepackage{lipsum}                    
\usepackage{mwe}                       
\usepackage{enumitem}
\usepackage{xurl} 
\usepackage{microtype}

\usepackage{mathptmx}                  

\onlineid{1022}

\vgtccategory{Empirical Study}

\vgtcinsertpkg

\usepackage{xcolor,soul} 

\newcommand{\text}[1]{\textbf{\textcolor[HTML]{56B4E9}{#1}}}
\newcommand{\gsq}[1]{\textbf{\textcolor[HTML]{999999}{#1}}}
\newcommand{\gh}[1]{\textbf{\textcolor[HTML]{424242}{#1}}}
\newcommand{\osq}[1]{\textbf{\textcolor[HTML]{E69F00}{#1}}}
\newcommand{\oh}[1]{\textbf{\textcolor[HTML]{F15A24}{#1}}}

\newcommand{\rev}[1]{\textcolor{black}{#1}}

\title{Visualizing in the Mind's Eye: \\Icon Design Shapes Mental Imagery of Fire Risks}

\author{Wen Xu\thanks{e-mail: \{xu.wen3, a.arunkumar, l.padilla\}@northeastern.edu} %
\and Anjana Arunkumar\footnotemark[1] %
\and Lace Padilla\footnotemark[1]}
\affiliation{\scriptsize Northeastern University}

\teaser{
  \centering
  \includegraphics[width=\linewidth]{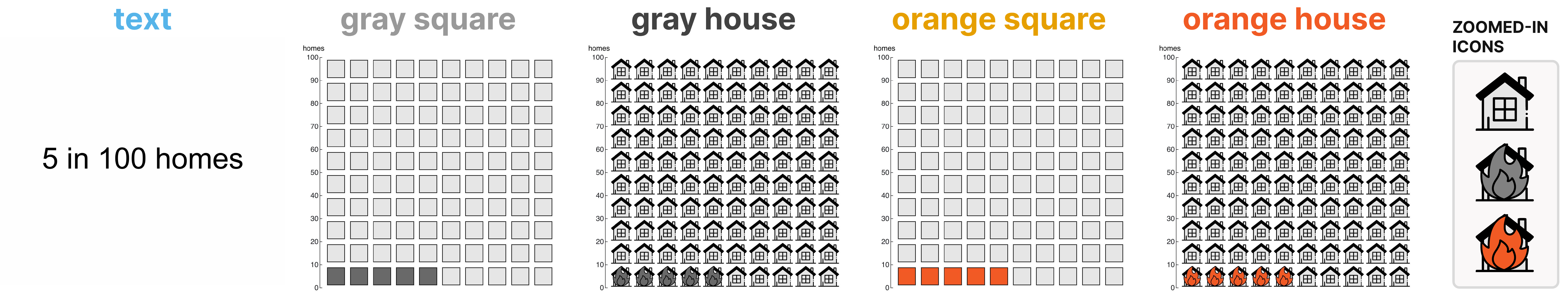}
  \caption{Risk information shown in the five conditions when probability = 5\%. Zoomed-in icons are shown on the right. }
  \label{fig:teaser}
}

\abstract{
We introduce mental imagery, or seeing images in the ``mind's eye," as a cognitive process that can be shaped by data visualization design and in turn impact decisions. We found in a preregistered study~(\textit{n}~=~400) that abstract geometric icons visualizing fire risk data evoked more mental images than concrete house-on-fire icons, and also produced more diverse and personalized mental images. Mediation analysis showed that increased mental imagery subsequently led to risk-averse decisions through evoking negative affect. These findings reveal a nuanced mechanism through which visualization concreteness influences decisions: concrete designs may actually suppress affect-driven behavior by restricting mental imagery.

} 


\begin{document}


\firstsection{Introduction}

\maketitle

Research on decision-making with data visualizations has shown that what viewers see can influence the decisions they make \cite{padilla2018decision}. 
Yet, a substantial body of research in adjacent fields suggests that decision-making is not only influenced by what viewers see on a page or screen, but also by what they see in their \textit{mind's eye}, which may or may not align with the external visual information \cite{adaval_doubt_2018, zaleskiewicz_decision_2023}. Forming visual \textit{mental imagery}, or images held in one's mind that are distinct from externally perceived visual information~\cite {adaval_doubt_2018}, is an important mechanism through which people simulate future events to inform decisions under uncertainty \cite{zaleskiewicz_decision_2023}. Despite the strong impact that visual information can have on priming, scaffolding, or restricting mental imagery \cite{adaval_doubt_2018, babin_effects_1997}, researchers have done little investigation on mental imagery as a process through which data visualization designs influence decisions. In this paper, we examine how visualization design shapes mental imagery about risk outcomes, which in turn alters decisions under risk. 

Mental imagery impacts decision-making under risk by enabling people to simulate possible outcomes and experience \textit{affect} (i.e., emotions) from those simulated scenarios \cite{zaleskiewicz_decision_2023}. Richer mental imagery may therefore lead to more risk-averse behavior by evoking stronger affect \cite{loewenstein2001risk}. For example, people tend to be risk-averse for incidents that more easily evoke vivid mental images and fear (e.g., terrorist acts) and under-prepare for risks associated with less vivid imagery (e.g., floods)~\cite{johnson_framing_1993}. 
We present a preregistered study\footnote{\url{https://osf.io/bfgru/}} to test the relationship between visualization design, mental imagery, affect, and risk decisions. We manipulate how \textit{concretely} visualizations represent wildfire risk outcomes, ranging from text to abstract icons to concrete icons depicting homes on fire (\cref{fig:teaser}). We find support for prior work's hypothesis that concreteness of visual representations influences the richness of mental imagery, which in turn evokes affect and impacts risk decisions. Specifically, the abstract icons \textit{increased} the quantity of viewers' mental imagery compared to concrete house-on-fire icons, which indirectly led to stronger negative affect and more risk-averse behavior. 

This work has the following contributions: 

\begin{itemize}[topsep=0pt, itemsep=0pt, parsep=0pt, partopsep=0pt]
    \item For researchers, we introduce mental imagery as a useful construct for understanding responses to risk visualizations, showing that visualization design can shape viewers' mental images of risk outcomes, which in turn impact risk decisions. 
    
    \item For visualization designers, we show that the influence of visualization concreteness on mental imagery can be nuanced and sometimes counterintuitive. We recommend using abstract icons when the designer aims to evoke more personal mental images from icon arrays, as their generic nature enables a wider range of mental imagery. 

\end{itemize}

\section{Related Work}

\textbf{Mental Imagery and Affect in Decision-Making. }Mental imagery, or forming an image of something in one's mind without it being physically present, has been extensively studied by cognitive scientists, psychologists, and philosophers~\cite{adaval_doubt_2018}. Contemporary research views imagery as a quasi-perceptual experience allowing us to perform mental simulations that inform decision-making~\cite{pearson_heterogeneity_2015, zaleskiewicz_decision_2023}. This is especially useful for decisions under uncertainty, as one can simulate different choice alternatives and their consequences to decide which is the most desirable~\cite{zaleskiewicz_decision_2023}. 

The role of mental imagery in decision-making is closely associated with emotions. 
The affect heuristic~\cite{slovic_affect_2007} and ``risk as feelings"~\cite{loewenstein2001risk} theories propose that people rely on rapid, automatic affective responses as shortcuts to make decisions under risk, and mental imagery of risk outcomes can be a source of such affect. Evidence from various decision-making domains supports this causal link~\cite{karlsson_causal_2023, lipkus_effects_2022, sobkow_affective_2016}. For example, 
instructing participants to engage in detailed mental imagery of environmental risk events led them to experience stronger negative affect and higher risk perception of climate change~\cite{karlsson_causal_2023}. 

Another body of research primarily in consumer psychology examined how visual information shapes mental imagery and subsequently drive\rev{s} behavior
, finding that using more concrete pictures in advertisements 
enhanced mental imagery of using the product and led to more positive attitudes~\cite{babin_effects_1997, petrova_fluency_2005}. Yet, other studies on the ``vividness effect," or the idea that more concrete information has a greater impact on attitude and behavior, have yielded mixed results~\cite{blonde_revealing_2016}. Similarly, the relationship between concrete visuals and mental imagery is not straightforward. Because mental imagery and visual perception draw on similar neural mechanisms~\cite{kosslyn_neural_2001}, external visual stimuli can interfere with mental imagery by competing for cognitive resources~\cite{unnava_interactive_1996}. Babin and Burns~\cite{babin_effects_1997} found that although concrete pictures produced more \textit{vivid} mental imagery, they did not increase the \textit{quantity} of mental imagery or its level of elaboration, possibly because ``everything is provided in the stimulus, which may actually stifle further mental imagery."

\textbf{Concrete Visualizations. }Some data visualizations depict the entities or topics behind them more concretely than others, using embellishments~\cite{bateman_useful_2010} or pictographs~\cite{burns_designing_2022} to vividly convey the semantic meaning behind the numbers. Researchers have found such concrete visualizations to facilitate memory~\cite{bateman_useful_2010}, engagement~\cite{haroz_isotype_2015}, and envisioning of the data topic~\cite{burns_designing_2022}, but have been inconclusive about whether they can further influence behavior. The anthropographics~\cite{morais_showing_2022} literature proposes that more realistic and specific visualizations of data about humans may promote pro-social behavior by evoking empathy. Yet, existing studies only found null or minimal effects of concrete anthropomorphic marks relative to abstract marks~\cite{boy_showing_2017, morais_can_2021}. Evidence from the risk communication field is also mixed, with some studies finding no effects of icon realism on risk perception~\cite{stone_effects_1997, blase_icon_2024} and others finding that concrete icons make viewers risk-averse~\cite{witteman_animated_2014, matzen_numerical_2023}. For example, participants were more likely to evacuate under a hypothetical fire risk if they saw icon arrays depicting houses on fire as opposed to squares~\cite{matzen_numerical_2023}. 

In this paper, we explore a possible explanation for these contradictory findings: concrete visualizations may only strongly influence those already concerned about the topic. Previous work has suggested that viewers' response to visualizations is dependent on their prior experiences, attitudes, and beliefs~\cite{peck_data_2019, markant_when_2023}. We hypothesize that those more concerned about a risk would react more intensely to concrete depictions of it, experiencing stronger mental imagery and affect and in turn make more risk-averse decisions. \rev{We ground our definition of concreteness in Construal Level Theory (CLT), which posits that more contextualized and less schematic representations are more concrete~\cite{trope2010construal}. CLT research shows that pictures are more concrete than text, color images are more concrete than grayscale images, and more photorealistic pictures are more concrete~\cite{trope2010construal, lee_monochrome_2014}. We thus manipulated graphical format, color, and icon realism as three aspects of concreteness in our stimuli. }

\section{Methods}

\subsection{Study Design} 
\textbf{Design. }This experiment adopts a mixed design. Participants were randomly assigned to one of five conditions: \text{text}, \gsq{gray square icons}, \gh{gray house icons}, \osq{orange square icons}, and \oh{orange house icons}~(\autoref{fig:teaser}). Each participant completed three trials with varying probability levels: 1\%, 5\%, and 9\%, presented in a random order.  

\textbf{Task and Stimuli. }Participants were asked to imagine being in a hypothetical scenario: ``\textit{Imagine that you own a home and will live in it for the next 30 years. There is a $X\%$ ($X$ = 1, 5, or 9) chance that a wildfire will damage your home in the next 30 years. If the wildfire happens, the damage to your properties will be \$80,000 (in today’s dollars). You do not have insurance to cover this loss. You have the option to invest in a one-time fire preparation for your home now, which would completely eliminate this risk for the next 30 years without needing further maintenance.}” Participants then entered the maximum amount of today’s dollars they would be willing to pay for the preparation in a text box, ranging between \$0-\$80,000. The probability, time frame, and damage amount were selected to reflect average fire risk levels in the US~\cite{firststreet, fire_damage}. The risk probability was presented as a natural frequency in an image under the description, as shown in~\cref{fig:teaser}. In the \text{text} condition, the image displays ``$X$ in 100 homes.” In other conditions, participants viewed a 10$\times$10 icon grid highlighting $X$ icons differing in color (gray vs. orange) and shape (squares vs. houses overlaid with fire).

\textbf{Procedures. }Participants completed the study via Qualtrics~\cite{qualtrics}. They read instructions about the tasks after informed consent approved by Northeastern University's IRB. If they passed an attention check, they were randomly assigned to one of the five conditions. They reported the amount they would be willing to pay for fire preparation 
and perception of the risk probability 
in three trials. They then retrospectively reported their mental imagery and affect when completing the task. Finally, they completed a manipulation check, a short graph literacy scale~\cite{okan_using_2019}, and demographic questions, including a 6-point item for gross annual household income \rev{from Qualtrics' certified demographic questions~\cite{qualtrics}.} 

\subsection{Measures}

\textbf{Mental Imagery. }We asked participants to describe any mental images that came to mind when they first viewed the risk information (any ``pictures" or ``videos" that they ``saw" in their mind's eye) in an open-ended question. We also used two 7-point Likert scales adapted from Miller et al.~\cite{miller_scale_2000} to measure the \textit{quantity} (3 items) and \textit{vividness} (7 items) of participants' mental images. 

\textbf{Affective Response. }We used the Berlin Emotional Responses to Risk Instrument~\cite{petrova_measuring_2023} to measure immediate affective response to the risk information. Participants were asked ``\textit{How did you feel when you read the information about the fire risks?}" and completed six 7-point Likert items for negative affect (\textit{anxious}, \textit{afraid}, \textit{worried}) and positive affect (\textit{assured}, \textit{hopeful}, \textit{relieved}). 

\textbf{Risk Preference. }Risk preference was calculated as participants' willingness to pay (WTP) for fire preparation minus the expected loss from the fire: $WTP - Damage \times Probability$, where $Damage = \$80,000$ and $Probability = 1\%$, $5\%$ or $9\%$. Higher values indicate higher risk aversion. Values $>0$ indicates risk-averse behavior, $<0$ indicates risk-seeking behavior, and $=0$ indicates risk neutrality.

\subsection{Hypotheses and Analysis}

\textbf{Preregistered Hypotheses and Analysis. }We hypothesized an interaction effect between presentation format and prior concern on mental imagery (\textbf{H1}), negative affect (\textbf{H2}), and risk preference (\textbf{H3}). Specifically, we hypothesized that participants with \textbf{higher prior concern} about fires will report higher quantity and vividness of mental imagery, experience stronger negative affect, and be more risk-averse in response to \textbf{more concrete} presentation formats. Across the sub-hypotheses of \textbf{H1-H3}, we made separate hypotheses for different design elements that we expect to make the risk information more concrete: \textbf{a)} format (Icons $>$ Text), \textbf{b)} color (Orange $>$ Gray), and \textbf{c)} icon type (House $>$ Square). \rev{We hypothesized these as isolated effects rather than a monotonic continuum, making no predictions about mixed conditions (e.g., \gh{gray house} vs. \osq{orange square}).}

We tested \textbf{H1} and \textbf{H2} with fixed-effects linear models using the \textit{glm()} function in the \textit{stats} package v. 4.6.0~\cite{R-stats} in R~\cite{R-base}: \( \mathit{Quantity / Vividness / Negative\ Affect}~\sim \mathit{Presentation\ Format} \times \mathit{Concern} \), where \textit{Concern} was mean-centered \rev{from a 1-10 Likert item}. \textbf{H3} was tested with a linear mixed-effects model using the $lmer()$ function in the $lme4$ package v. 1.1.37~\cite{lme4}: \( \mathit{Risk\ Preference} \sim \mathit{Presentation\ Format} \times \mathit{Concern} + \mathit{Probability} + \mathit{Income} + \mathit{Graph\ Literacy} + \mathit{(1 | ID)}\). \textit{Probability} was the within-subjects probability level serving as a covariate. \textit{Income} and \textit{Graph Literacy} were also covariates and mean-centered. \(\mathit{(1 | ID)}\) was a random intercept for each participant. 

\textbf{Conceptual Diversity and Personal Relevance of Mental Imagery. }As a measure of the conceptual diversity of mental images, we conducted inductive qualitative coding of the conceptual categories appearing in participants' self-reported mental images and calculated a count of distinct categories for each participant. We also coded whether the mental image had a personally relevant subject (i.e., if participants imagined the fire impacting themselves and/or their real-life acquaintances). The codebook was iteratively developed by two coders blind to the experimental conditions. The first coder developed an initial codebook after familiarizing themselves with all responses. The two coders then independently coded a random 15\% of the dataset using the codebook, resolved discrepancies through discussion, and refined the codebook. They then independently coded the rest of the dataset, reaching an inter-coder reliability score (Cohen's $\kappa$) of 0.75 and resolved discrepancies. The final codebook for conceptual diversity had a hierarchical structure with 27 codes under four themes: entities (e.g., fire, home, victims), events (e.g., firefighting, evacuation), concepts (e.g., damage, loss), and settings (e.g., place name, media). The codebook is available in the supplemental materials.

\textbf{Serial Mediation Analysis. }To test our hypothesized causal chain (\textit{Presentation Format} $\rightarrow$ \textit{Mental Imagery} $\rightarrow$ \textit{Negative Affect} $\rightarrow$ \textit{Risk Preference})
, we tested a multilevel serial mediation model using Structural Equation Modeling (SEM) 
with the \textit{lavaan} package v. 0.6.21~\cite{rosseel_lavaan_2012} in R. At the within-subject level, risk preference was modeled as a function of trial-level probability. At the between-subject level, we specified a serial mediation model with presentation format and mean-centered concern as primary predictors, imagery quantity as the first mediator (M1), and negative affect as the second mediator (M2). For both presentation format and concern, the model simultaneously estimated the direct effect on risk preference, as well as three mediation pathways: a specific indirect effect through M1 alone, a specific indirect effect through M2 alone, and a serial indirect effect through M1 and then M2.

\subsection{Participants} 

We used the 
the $simr$ package~\cite{simr} in R to estimate the sample size needed to detect our hypothesized interaction effect with an 80\% power, 78 participants per group. We rounded up to 90 per group, or \textit{n} = 450 in total, to account for preregistered exclusions: failing an attention check or a manipulation check confirming accurate perception of stimuli color.

We recruited adults on Prolific~\cite{prolific} who were based in the US, fluent in English, had an approval rate of over 80\%, completed the study on a desktop, and had not participated in previous pilots. As our pilots indicated that the majority of the Prolific population had relatively low levels of concern about fire risks, we conducted a pre-screening survey to select a balanced sample of participants with low versus high levels of fire concern for the main study. The survey included a 10-point Likert item for concern about wildfire threat to one's home adapted from \rev{the Concern dimension of }the Householder Bushfire Risk Perception Scale~\cite{hall_conceptualising_2022} and four binary-choice questions asking whether the participant or someone close to them had been impacted by a wildfire. 
We randomly selected a subsample of screening survey participants to invite to the main study 48 hours later. The invited sample was stratified by fire concern, where 50\% of invited participants reported low concern (ratings of 1–5) and 50\% reported higher concern (ratings of 6–10).

We sent the screening survey to 1,014 participants on Prolific, invited 608 back to the main study, and terminated the study when we received at least 90 responses in every condition. The final sample size was \textit{n =} 463 and \textit{n =} 400 after preregistered exclusions, of which 203 identified as female, 191 as male, and six as non-binary. Participants had a mean age of 38 (\textit{SD} = 12), graph literacy of 2.36 out of 4 (\textit{SD} = 1.04), \rev{and fire concern of 5.22 out of 10 (\textit{SD} = 2.81).} The study took 11 minutes 28 seconds on average at \$12/hour.

\section{Results}

\subsection{Preregistered Analysis}

\begin{figure}[t]
    \centering
    \includegraphics[width=0.9\columnwidth, alt={Probability distributions of mental imagery quantity (1-7) for each presentation format. Participants who viewed gray squares reported higher quantity of mental imagery than orange houses, p < .01. } ]{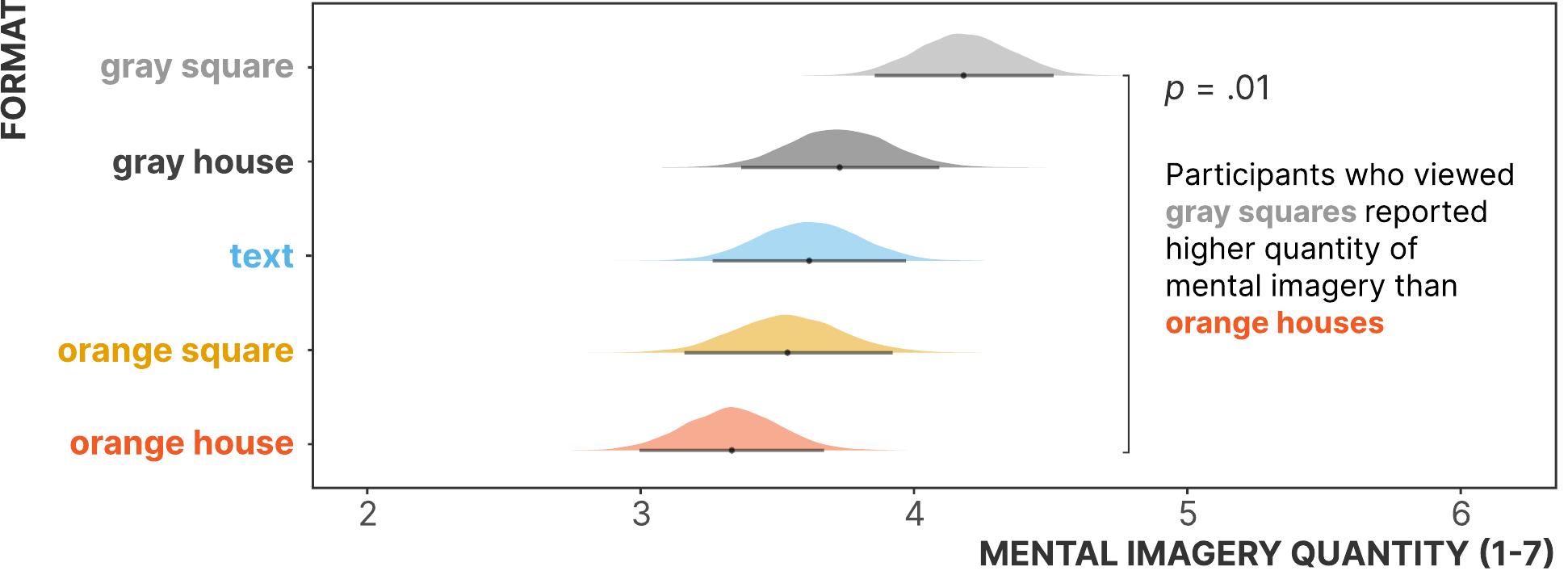}
    \caption{Mental imagery quantity across presentation formats. Distributions are based on bootstrapped 95\% CIs, which the horizontal bars represent. }
    \label{fig:imagery-quantity}
\end{figure}

\textbf{Mental Imagery. }We found no significant interaction between presentation format and concern on mental imagery \textit{\textbf{quantity}}. A likelihood ratio test showed that the preregistered interaction model did not significantly improve model fit compared to a simpler model with only main effects (\( \mathit{Presentation\ Format}\ + \mathit{Concern}\)), \textit{F}(4, 390) = 1.78, \textit{p} = .13. We therefore used the parsimonious main-effects model for subsequent analysis. This model showed a significant main effect of presentation format on mental imagery quantity, $\chi^2(4) = 11.43$, \textit{p} = .02. As shown in \cref{fig:imagery-quantity}, post hoc pairwise comparisons with Tukey adjustment found that participants who viewed \gsq{gray squares} reported higher \textit{quantity} of mental imagery than \oh{orange houses}, \textit{b} = 0.82, 
\textit{t}(394) = 3.21, \textit{p} = .01. Prior concern also increased imagery quantity, \textit{b} = 0.13, 
\textit{t}(394) = 4.64, \textit{p} = .0001.

There was an interaction between presentation format and concern on mental imagery \textit{\textbf{vividness}}, $\chi^2(4) = 12.52$, \textit{p} = .01. Tukey-adjusted pairwise comparisons showed that for participants with a high level (\textit{Mean} + 1\textit{SD}) of prior concern, the \oh{orange house} icon arrays evoked \textit{less vivid} mental imagery than \text{text} (\textit{b} = -1.03, \textit{t}(390) = -2.83, \textit{p} = .04), \gsq{gray square} (\textit{b} = -1.07, \textit{t}(390) = -2.92, \textit{p} = .03), and \gh{gray house} (\textit{b} = -1.24, \textit{t}(390) = -3.47, \textit{p} = .01).

\textbf{Negative Affect. }There was no significant interaction between presentation format and concern on negative affect. We again compared the interaction model with a main-effects model and found no significant improvement of fit, \textit{F}(4, 390) = 0.64, \textit{p} = .63. The main-effects model showed that prior concern increased negative affect from viewing the stimuli, \textit{b} = 0.20, \textit{t}(394) = 6.77, \textit{p} $<$ .0001. 

\textbf{Risk Preference. }The preregistered model found no interactions and did not improve fit compared to a main-effects model, $\chi^2(4) = 1.75$, \textit{p} = .78. Prior concern increased risk aversion in the main-effects model, \textit{b} = 750.87, \textit{t}(392) = 3.49, \textit{p} = .0005.

\subsection{Exploratory Analysis}

\textbf{Abstract Icons Increased Diversity and Personal Relevance of Mental Imagery. }Our preregistered analysis reveals unexpected insights: the more abstract \gsq{gray squares} evoked \textit{higher quantity} of mental imagery than the most concrete \oh{orange houses}, contrary to our hypothesis that concrete visualizations increase mental imagery. Similarly, concerned individuals experienced \textit{less vivid} mental imagery when viewing \oh{orange houses} than more abstract formats. 
One likely explanation is that abstract icons allow for more diverse and more personalized mental images than concrete icons. The \oh{orange houses} may have limited mental images to a specific burning house, whereas one can more easily imagine a wider range of concepts with \gsq{gray squares}, including people, pets, or detailed personal memories. 

Consistent with our reasoning, participants who viewed \gsq{gray squares} had the highest number of distinct concepts in their mental imagery descriptions on average ($\bar{x} = 3.20$), and those who viewed \oh{orange houses} had the lowest ($\bar{x} = 2.23$). One participant specifically described seeing the \gsq{gray squares} as people: ``\textit{...a graph of 100 people where some people are in one color when they have a characteristic while the rest is blank outline of a person.}" There was a similar trend with personal relevance: the \gsq{gray square} group had the highest proportion of participants imagining personally relevant fire consequences ($\bar{x} = 49\%$), while \oh{orange house} ($\bar{x} = 30\%$) and \osq{orange square} ($\bar{x} = 29\%$) had the lowest.

\begin{figure}[!htb]
    \centering
    \includegraphics[width=\columnwidth, alt={}]{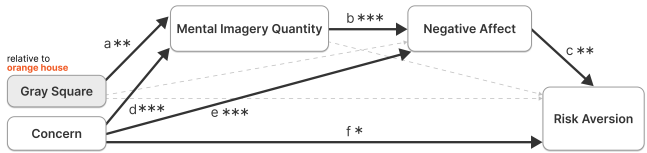}
    \caption{Conceptual diagram of the serial mediation model. * denotes \textit{p} $<$ .05, ** denotes \textit{p} $<$ .01, and *** denotes \textit{p} $<$ .001. Dashed lines indicate non-significant paths. The model also includes all conditions (dummy-coded with \oh{orange house} as referent) as predictors of imagery quantity, affect, and risk aversion at the between-subjects level and probability as a predictor of risk aversion at the within-subjects level. They are omitted from the diagram to reduce clutter. }
    \label{fig:mediation-fig}
\end{figure}

\textbf{Indirect Effect of Abstract Icons on Risk Preference Via Mediation of Imagery Quantity and Affect. }Although \textbf{H2} and \textbf{H3} found no main effects of conditions on negative affect and risk preference, exploratory mixed-effects models\footnote{Same structure as \textbf{H1}'s model, replacing \(\mathit{Presentation\ Format} \times \mathit{Concern}\) with \textit{Imagery Quantity} or \textit{Negative Affect}} showed that mental imagery quantity (\textit{b} = 1163.61, \textit{t}(396) = 3.12, \textit{p} = .002) and negative affect (\textit{b} = 1518.62, \textit{t}(396) = 4.41, \textit{p} $<$ .0001) both strongly predicted risk aversion. This motivated us to investigate whether \gsq{gray squares} influence risk behavior through a more specific mechanism, or an indirect-only effect~\cite{zhao_reconsidering_2010}. We used a serial mediation model to test if \gsq{gray squares} indirectly influenced risk preference through the causal pathway proposed by existing theories~\cite{loewenstein2001risk, zaleskiewicz_decision_2023}. Specifically, we tested if visual stimuli first impact imagery quantity, imagery then evokes negative affect, and affect ultimately increases risk aversion (pathways $a \to b \to c$ shown in~\cref{fig:mediation-fig}). \rev{We focus on quantity in this analysis because vividness did not significantly predict risk aversion in exploratory mixed-effects models. }

We found a significant serial indirect effect of \gsq{gray squares} increasing risk aversion relative to \oh{orange houses}, operating sequentially through imagery quantity and negative affect, $\beta$ = 0.01, \textit{p} = .04 ($a \to b \to c$ in~\cref{fig:mediation-fig}). \gsq{Gray squares} increased imagery quantity, $\beta$ = 0.20, \textit{p} = .001 (path $a$), which heightened negative affect, $\beta$ = 0.35, \textit{p} $<$ .001 ($b$), which in turn increased risk aversion, $\beta$ = 0.16, \textit{p} = .005 ($c$). Consistent with an indirect-only mediation process, neither the direct effect of \gsq{gray squares} on risk preference ($\beta$ = 0.10, \textit{p} = .12) nor its total effect ($\beta$ = 0.13, \textit{p} = .06) were significant. In addition, prior concern had a significant total effect on risk aversion, $\beta$ = 0.20, \textit{p} $<$ .001, comprising of an indirect effect through negative affect, $\beta$ = 0.04, \textit{p} = .01 ($e \to c$), a serial indirect effect through imagery quantity and affect, $\beta$ = 0.01, \textit{p} = .02 ($d \to b \to c$), and a direct effect, $\beta$ = 0.13, \textit{p} = .02 ($f$). We also tested an alternative model reversing the order of the mediators to evaluate the possibility of affect leading to more imagery. This reverse pathway did not yield significant serial indirect effects for condition or concern, suggesting that imagery preceded affect in the mediation sequence. 

\section{Discussion and Conclusion}

\textbf{Visualization Concreteness and Mental Imagery. }Our findings showed a consistent difference between the most abstract (\gsq{gray square}) and most concrete (\oh{orange house}) icon designs tested: \oh{orange houses} reduced the quantity of mental imagery overall and vividness for concerned individuals. These findings challenge the intuitive hypothesis that more concrete visual information leads to richer mental imagery. Qualitative analysis suggests that the concrete icons restricted the range of concepts that can appear in mental images and limited personally relevant imagery, possibly because of their specificity. It is intriguing that \text{text}, which we assumed to be more abstract than any graphical format, did not produce more mental images than icon arrays. It is possible that the icon array format provides a visual scaffold for imagery that is not present in \text{text}, as some participants described using the icon grid as a basis to imagine impacted entities (e.g., burning or damaged houses and people). Given that scaffold, abstract icons may have allowed for a wider range of imagery than concrete ones.

\textbf{Imagery, Affect, and Risk Decisions. }The serial indirect effect partially supports the causal relationship hypothesized by the ``risk as feelings" theory. Specifically, information concreteness shapes risk decisions by increasing mental imagery that evokes affect, although the effect of concreteness was surprisingly in the opposite direction. However, because the total effect of \gsq{gray squares} on risk preference was not significant (\textit{p} = .06), we caution against concluding that visualization concreteness meaningfully altered risk decisions and encourage future work to explore situations in which it does, potentially with more concrete designs such as photographs.

\textbf{The Role of Prior Concern. }We found that prior concern about fire risks can have a more profound effect on behavior than visualization design. It is the strongest predictor of not only mental imagery and affect, but also risk-averse decisions. We also reveal that responses to concrete visualization design can depend on one's preexisting concern about the topic. Specifically, concrete icons led to less vivid mental imagery only for highly concerned individuals. Echoing prior work~\cite{peck_data_2019, markant_when_2023}, this highlights the need for visualization designs to account for the audience's predispositions, not only individual differences in ability (e.g., graph literacy), but also preexisting affect and attitude about the broader data context. 

\textbf{Limitations. }We elicited mental imagery with an asynchronous written measure. Given the spontaneous and visual nature of mental imagery, a real-time think-aloud procedure or a drawing task may be more reliable. The imagery and affect measures were retrospective to prevent instructions from priming decisions, \rev{which limits our ability to confirm the temporal sequence of variables. While existing theory~\cite{loewenstein2001risk} and the lack of significance in reverse models justify our mediation sequence, the retrospective measures remain subject to memory effects. }
\rev{In addition, varying icon color may have altered salience or semantic associations, rather than just concreteness. Although the consistent difference between \gsq{gray squares} and \oh{orange houses} suggest combined effects of color \textit{and} icon type and thus likely concreteness as a whole, future work is needed to evaluate whether confounding constructs provide alternative explanations for our findings. }

\textbf{Design Implications and Conclusion. }This study reveals a counterintuitive relationship between visualization concreteness and viewers' mental imagery, and demonstrates how mental imagery can shape risk decisions through evoking affect. Given the substantial impact of mental imagery on emotional and behavioral responses to visualizations, we encourage designers to consider it as a design objective when creating visualizations with real-world decision-making implications. Perhaps contrary to design intuitions, we specifically recommend designers to use abstract, generic marks over concrete, specific ones when they aim to enable more diverse or personally relevant visual imaginations about data. Prior concern of the target audience is also important---a viewer who is already concerned about the data topic can be particularly limited by concrete designs. Ultimately, we highlight mental imagery as an underexplored mechanism that shapes reactions to visualizations. By designing for the ``mind's eye," researchers and practitioners can uncover new ways to impact decisions with data.

\section*{Supplemental Materials}
\label{sec:supplemental_materials}
Supplemental materials available at~\url{https://osf.io/ugwmh/}.

\renewcommand{\UrlFont}{\ttfamily\small} 
\bibliographystyle{abbrv}

\bibliography{template}

\end{document}